\documentclass[12pt]{article}

\input{epsf}
\usepackage[dvips]{graphicx}
\usepackage{cite}
\def\1ad{\mbox{\normalsize $^1$}}
\def\2ad{\mbox{\normalsize $^2$}}
\def\3ad{\mbox{\normalsize $^3$}}
\def\4ad{\mbox{\normalsize $^4$}}

\def\5ad{\mbox{\normalsize $^5$}}
\def\6ad{\mbox{\normalsize $^6$}}
\def\7ad{\mbox{\normalsize $^7$}}
\def\8ad{\mbox{\normalsize $^8$}}

\def\dj{\hbox{d\kern-0.347em \vrule width 0.3em height 1.252ex depth
-1.21ex \kern 0.051em}}

\newcommand{\be}{\begin{equation}}
\newcommand{\ee}{\end{equation}}
\newcommand{\ben}{\begin{equation*}}
\newcommand{\een}{\end{equation*}}
\newcommand{\ba}{\begin{eqnarray}}
\newcommand{\ea}{\end{eqnarray}}
\newcommand{\ban}{\begin{eqnarray*}}
\newcommand{\ean}{\end{eqnarray*}}
\newcommand{\brr}{\begin{array}}
\newcommand{\err}{\end{array}}
\newcommand{\bc}{\begin{center}}
\newcommand{\ec}{\end{center}}

\newcommand{\bea}{\begin{eqnarray}}
\newcommand{\eea}{\end{eqnarray}}
\newcommand{\bean}{\begin{eqnarray*}}
\newcommand{\eean}{\end{eqnarray*}}

\newcommand\lsim{\mathrel{\rlap{\lower4pt\hbox{\hskip1pt$\sim$}}
    \raise1pt\hbox{$<$}}}
\newcommand\gsim{\mathrel{\rlap{\lower4pt\hbox{\hskip1pt$\sim$}}
    \raise1pt\hbox{$>$}}}

\newcommand{\centeron}[2]{{\setbox0=\hbox{#1}\setbox1=\hbox{#2}\ifdim
                             \wd1>\wd0\kern.5\wd1\kern-.5\wd0\fi \copy0
                             \kern-.5\wd0\kern-.5\wd1\copy1\ifdim\wd0>\wd1
                             \kern.5\wd0\kern-.5\wd1\fi}}
\newcommand{\ltap}{\>\centeron{\raise.35ex\hbox{$<$}}
                     {\lower.65ex\hbox{$\sim$}}\>}
\newcommand{\gtap}{\>\centeron{\raise.35ex\hbox{$>$}}
                     {\lower.65ex\hbox{$\sim$}}\>}

\begin{document} 

\setcounter{page}{0}
\thispagestyle{empty}

\begin{flushright}
CERN-PH-TH/2006-125\\
hep-ph/0607107
\end{flushright}

\vskip 8pt

\begin{center}
{\bf \Large {
Gravitational Waves from Phase Transitions\\ [0.25cm]  
 at the Electroweak Scale and Beyond
}}
\end{center}

\vskip 10pt

\begin{center}
{\large Christophe Grojean $^{a,b}$ and  G\'eraldine Servant $^{a,b}$ }
\end{center}

\vskip 20pt

\begin{center}
\centerline{$^{a}${\it CERN, Theory Division, CH-1211 Geneva 23, Switzerland}}
\centerline{$^{b}${\it Service de Physique Th\'eorique, CEA Saclay, F91191 Gif--sur--Yvette,
France}}
\vskip .3cm
\centerline{\tt  christophe.grojean@cern.ch,geraldine.servant@cern.ch}
\end{center}

\vskip 13pt

\begin{abstract}
\vskip 3pt
\noindent
If there was a first order phase transition in the early universe, there should be an associated stochastic background of gravitational waves. In this paper, we point out that the characteristic frequency of the spectrum due to  phase transitions which took place in the temperature range 
100~GeV -- ${10^7}$~GeV is precisely in the window that will be probed by the second generation of space-based interferometers such as the Big Bang Observer (BBO). Taking into account the astrophysical foreground, we determine the type of phase transitions which could be detected either at LISA, LIGO or BBO, in terms of the  amount of supercooling and the duration of the phase transition that are needed. Those two quantities can be calculated for any given effective scalar potential describing the phase transition. 
In particular, the new models of electroweak symmetry breaking which have been proposed in the last few  years typically have a different Higgs potential from the Standard Model. 
They could lead to a gravitational wave signature in the milli-Hertz frequency, which is precisely the peak sensitivity of LISA.
We also show that
the signal coming from phase transitions taking place at T$\sim $ 1--100 TeV could entirely screen the relic  gravitational  wave signal expected from standard inflationary models.

\end{abstract}

\vskip 13pt
\newpage
\section{Introduction}

Direct detection of gravitational radiation will hopefully soon become a reality, with the operation of the first generation of 
 interferometers such as (kilometer-scale, ground-based) LIGO~\cite{LIGOURL} and VIRGO~\cite{VIRGOURL} and (million-kilometer-scale, space-based) LISA~\cite{LISAURL}. Those instruments will allow us to probe gravitational waves produced by astrophysical objects at relatively low red-shift (black hole binaries, neutron star binaries, white dwarf binaries, supernovae, pulsars...)
In addition, gravitational waves (GW) can provide information about particle physics at unexplored high energies.
The weakness of the interaction with matter is a major obstacle for detection of gravitational waves but it also has the virtue that the information they carry about the state of the universe at the moment of their production has been unaltered. 
They are precious information on the very mechanism that produced them.
GW can be produced by core collapse of supernovae, first-order phase transitions, vibration of cosmic strings, preheating, dynamics of extra dimensions\ldots

Among those well-motivated but hypothetical cosmological sources of GW, there is at least one that we are convinced exists:   the GW produced during inflation. This signal is expected to be very tiny. Quantum fluctuations in the inflaton field during inflation leaves behind a residue in density perturbations observed in the Cosmic Microwave Background (CMB). They also lead to a background of GW whose properties couple with those of density fluctuations. As the CMB anisotropies are affected by GW, the WMAP constraint on the energy scale of inflation 
fixes a bound on the size of the GW signal due to inflation $\Omega_{GW} h^2 \lsim 10^{-15} -10^{-14}$~\cite{Smith:2005mm}. This is several orders of magnitude below the best sensitivity of the first-generation of interferometers.
However, attempts to detect this relic primordial background are very strongly motivated. This is a main goal of the second generation of space interferometers, in particular, the Big Bang Observer (BBO)~\cite{BBOURL}, the follow-on mission to LISA,  which would become a reality within twenty or thirty years (by comparison, LISA, if funded, should be operational by 2014).

The present work focusses instead on the detectability of GW from first-order phase transitions. 
The corresponding relic GW background encodes useful information on these major symmetry-breaking events which took place in the early universe.
In contrast with the inflationary spectrum, the spectrum is not flat, with a characteristic peak related to the temperature at which the phase transition (PT) took place. This  signal can actually be higher by several orders of magnitude than the signal expected from inflation and in some cases can entirely screen it. 
One symmetry-breaking event  which for sure took place in the early universe is electroweak (EW) symmetry breaking. What we do not know yet is whether it was a  first order phase transition, in which case it proceeded through  nucleation of bubbles resulting in a large departure from thermal equilibrium. Bubble collision and associated motions in the primordial plasma are sources of gravitational waves. 
The characteristic frequency  of the signal is close to the Hubble frequency at the time of the transition $H(T_{EW}) \sim 10^{-14}$~GeV. Once redshifted to today, this corresponds to mHz frequencies, 
which is precisely the frequency band that  LISA is sensitive to. It is therefore very exciting that LISA could help providing information on the EW scale, in particular on the nature of 
the EWPT.

The GW spectrum resulting from first order PT was computed 
 in the early nineties~\cite{Kosowsky:1992rz,Kosowsky:1991ua,Kosowsky:1992vn,Kamionkowski:1993fg} but this topic has not received much subsequent attention, as it was found out that there is no first order EWPT in the Standard Model given the experimental bound on the Higgs mass~\cite{lattice}.
 It was realized ten years after the original calculation of~\cite{Kosowsky:1992rz,Kosowsky:1991ua,Kosowsky:1992vn,Kamionkowski:1993fg} that turbulence in the plasma could be a significant source of GW in addition to bubble collisions~\cite{Kosowsky:2001xp, Dolgov:2002ra}. Subsequently, the authors of~\cite{Apreda:2001us} studied the GW signal due to a first order EWPT in the Minimal Supersymmetric Standard Model (MSSM) and its NMSSM extension. Finally, Nicolis~\cite{Nicolis:2003tg} did a model-independent analysis for the detectability of GW with LISA. 
 
 We believe that it is time to revisit this question for two reasons: The nature of the EWPT will start to be probed experimentally at the LHC. 
 Indeed, it depends essentially on the Higgs sector of the theory or any alternative dynamics for EW symmetry breaking. 
 In the last few years, new models of EW symmetry breaking have been suggested (little higgs, gauge-higgs unification, composite higgs, higgsless models ...) and the nature (smooth cross-over or first-order) of the EWPT in these new frameworks remains unknown. Second, the technology for gravitational wave detectors has made advances~\cite{Ballantini:2005am} and we think it is timely to  redo a model-independent analysis not only for LISA but also other devices. 

LIGO is sensitive to much higher  frequencies (from a few Hz to a few hundreds of Hz) thus it is in principle sensitive to phase transitions  which took place at much earlier epochs. For instance, we will show that if there was a very strong first order PT at temperatures of order $10^7$~GeV, the ultimate stage of LIGO (LIGO-III, correlated) could detect the corresponding peak (as already pointed out in~\cite{Kosowsky:1992rz}).
The second generation of interferometers will be able to say much about the possible existence of early universe first-order phase  transitions. Indeed, BBO will be sensitive to signals from PT which would have taken place in the temperature range $T\sim 100$~GeV-- $10^7$~GeV, even if not necessarily exceptionnally strong. 

In this paper, we start with some generalities on stochastic GW backgrounds including the astrophysical background. We also recap what would be the observable redshifted signal we would observe after the GW have propagated forward from the phase transition until today. Section 3 reviews the key formulae used in the theoretical predictions of the GW spectrum due to first-order phase transitions. There is nothing new in this part. However, this formalism had so far only been exploited to study the detectability at LISA of GW due to a first-order electroweak phase transition. In section 4, we apply it to any other phase transitions taking place at higher temperatures and compare them with the sensitivities of not only LISA but also LIGO and BBO.
Predictions can be presented in a model-independent way as a function of two quantities, namely $\alpha$ ($\sim$ latent heat) and $\beta^{-1}$ ($\sim$ duration of the phase transition), which can be computed for any given effective scalar potential describing the transition.
For each temperature, we identify which values of $\alpha$ and $\beta$ lead to an observable signal.  
Particle model builders can then test their favourite scalar potential by computing its corresponding values of $\alpha$ and $\beta$ and see whether it can give rise to a detectable GW signal.
We comment on some specific examples of particle physics models.

\section{Astrophysical versus cosmological GW background}

Stochastic backgrounds are random gravitational waves arising from the incoherent superposition of a large number of independent, uncorrelated sources that cannot be resolved individually. They are discussed in terms of their contribution to the universe's energy density, over some frequency band:
\be
\Omega_{GW}(f)=\frac{1}{\rho_{\mbox{\tiny crit}}}\frac{d \rho_{GW}}{d \ln f}
\ee
By their very nature, stochastic GW are indistinguishable from the detector noise. Ground-based detectors look for them by coordinated measurements (comparing outputs of multiple detectors to find sources of correlated noise) while LISA can extract the instrumental noise power by combining the signals from its three spacecrafts. For technical aspects related to the detection of a gravitational wave stochastic background, see  Ref.~\cite{Maggiore:1999vm}.

\subsection{Astrophysical foreground}
\label{astro_back}
Searching for GW waves of cosmological origin is an ambitious goal. There is a huge foreground due to astrophysical sources 
which in principle makes detection impractical.
Once the signals from every merging neutron star and stellar mass black holes have been identified and substracted,
the primary sources of foreground signals are galactic and extragalactic binaries.
The galactic background produced by binary stars in the Milky Way is many times larger in amplitude than both the extragalactic foreground and LISA's design sensitivity. However, it can be substracted because of its anisotropy, being mostly concentrated in the galactic plane.
Irreducible background comes from extragalactic binary stars and is dominated by emission from white dwarves (WD) pairs. The corresponding GW spectrum was estimated in Ref.~\cite{Farmer:2003pa} where limits are placed on the minimum and maximum expected background signals. Ref.~\cite{Farmer:2003pa} points out that at frequencies $f \lsim 50$ mHz, there will be too many individual WD--WD sources contributing in each resolution element to be completely resolved and substracted  source by source by missions with plausible lifetimes. However, much of the flux comes from relatively nearby sources, and the WD--WD numbers drop rapidly above 50 mHz. Thus it may be possible for future missions more sensitive than LISA to substract this background at high frequencies~\cite{Farmer:2003pa}.
In our figures, we plot  this background coming from unresolved compact white dwarf binaries assuming that it can be removed at frequencies above 50 mHz. 
At higher frequencies, the dominant foreground GW sources are inspiralling neutron star-neutron star, neutron star-black hole and black hole-black hole binaries. These have to be individually identified and substracted. This problem is discussed in~\cite{Cutler:2005qq} and in our BBO  detectability analysis we optimistically assume that this foreground can be substracted.

\subsection{Relic background from cosmological processes}

The GW background due to early universe events is stochastic as the signal comes from the superposition of incoherent sources originating from a huge number of different horizon volumes.
For instance, the size of the horizon at the time of the electroweak phase transition was much
 smaller than today ${(10^{-14} ~\mbox{GeV})}^{-1}$, corresponding to a tiny fraction of degree on the sky today.
Even if we assumed that there were two bubbles per horizon volume
(actually there would typically be several hundreds of them as we will see later),
we would be unable to resolve the signal coming from their collision.
The signal comes from bubble collisions which
took place in many independent universes.

We work in a standard Friedman--Robertson--Walker (FRW) cosmology, $a(t)$ is the cosmological scale factor. At the energy scales considered, we assume a radiation-dominated era.
Gravity waves produced at $T_*$ with a characteristic frequency $f_*$ propagate until today without interacting. Their energy density redshifts as $a^{-4}$ and their frequency as $a^{-1}$. The characteristic frequency we observe today is 
\be
f=f_* \frac{a_*}{a_0}=f_* \left( \frac{g_{s0}}{g_{s*}}\right)^{1/3}\frac{T_0}{T_*}
\ee
where we used the adiabaticity of the expansion of the universe (meaning that the entropy per comoving volume $S\propto a^3 g_s(T) T^3$ remains constant).
\be
g_s(T)= \sum_{i=\textit{\small bosons}}g_i\left(\frac{T_i}{T}\right)^3 
+\frac{7}{8}\sum_{i=\textit{\small fermions}}g_i \left(\frac{T_i}{T}\right)^3
\label{gs_definition}
\ee
$g_i$ counts the internal degrees of freedom of the $i$-th particle and the sum is over relativistic species.
Today, $g_s(T_0)\simeq 3.91$ (assuming three neutrino species) and $T_0= 2.725 K=2.348\times 10^{-13}$~GeV. It is convenient to express the frequency in terms of the Hubble frequency at the time of GW production:
\be
\label{characteristic_frequency}
f\approx 6 \times 10^{-3} \mbox{mHz}    \left(\frac{g_*}{100}\right)^{1/6} \frac{T_*}{100\mbox{ GeV}} \ \frac{f_*}{H_*}
\ee
The remarkable fact is that for $T\sim 100$~GeV and $f_*/H_* \sim 10^2$, (as expected for weak scale processes as will be explained below), the peak frequency  of the GW spectrum is in the milliHertz, just in the band of LISA.

The fraction of the critical energy density in gravity waves today is
\be
\Omega_{GW}=\frac{\rho_{GW}}{\rho_c}= \Omega_{GW*}\left(\frac{a_*}{a_0}\right)^4\left(\frac{H_*}{H_0}\right)^2\simeq 1.67\times 10^{-5} h^{-2}\left(\frac{100}{g_*}\right)^{1/3} {\Omega}_{GW*}
\ee
where we used
\be
{\rho}_{GW}={\rho}_{GW*}\left(\frac{a_*}{a_0}\right)^4 \ , \ \ \rho_c=\rho_{c*}\frac{H_0^2}{H_*^2} \  \mbox{and} \ H_0=2.1332\times h\times 10^{-42}~ \mbox{GeV}
\ee
$g_*$ is the number of relativistic degrees of freedom at $T_*$ which enters the definition of the energy density and not the entropy (it is given by Eq.~(\ref{gs_definition})  where the cubic power is replaced by a quartic power). $\Omega_{GW*}$ is the fraction of energy density of the universe at the time of the transition which is in gravitational waves. The peak sensitivity of LISA would correspond
 to detect $\Omega_{GW}h^2  \sim 10^{-11}$. This means that to detect a signal at LISA, we need $\Omega_{GW*}\gsim 10^{-6}$ while at BBO, we can probe smaller fractions, $\Omega_{GW*}  \sim 10^{-12}-10^{-9}$.

The remaining task is to estimate $f_*$ and $\Omega_*$.
 Theoretical predictions of relic GW backgrounds are subject to large uncertainties which depend on the cosmological mechanism. However, we can get a reasonable estimate of the characteristic frequency, the form of the spectrum and the typical intensity. We will now review the main results in the case of GW produced during first order phase transitions.

\section{GW from first-order phase transitions}

Phase transitions are commonly described by the effective potential of the scalar field (either elementary or composite) responsible for the dynamics. 
First-order phase transitions are triggered if there exists a temperature at which a barrier separates two degenerate minima. This happens for instance if there are negative cubic or quartic self couplings for the scalar field. 
In this case, phase transitions proceed via nucleation of bubbles of the low-temperature phase within the high-temperature phase.
Bubble nucleation occurs through quantum tunneling and thermal fluctuations. 
These bubbles then expand and merge, leaving the universe in the low-temperature phase (commonly the broken-symmetry phase).
As a bubble expands, part of the liberated latent heat raises the plasma temperature while the other part is converted into kinetic energy of the bubble wall and bulk motions of the fluid. 
Because of its spherical symmetry, a single expanding bubble produces no gravity waves. Only after bubble collisions destroy the spherical symmetry is gravitational radiation emitted.
High velocities and large energy densities provide the necessary conditions for producing gravitational radiation.
There are two sources of gravitational waves: the actual collision of bubbles and the turbulence in the plasma due to bubble motion.
The resulting spectrum of gravitational waves has been studied in details in~\cite{Kosowsky:1992rz,Kosowsky:1991ua,Kosowsky:1992vn,Kamionkowski:1993fg,Kosowsky:2001xp,Dolgov:2002ra,Apreda:2001us,Nicolis:2003tg}. The turbulence spectrum was recently revisited in Ref.~\cite{Caprini:2006jb}. Re-examination of the bubble collision spectrum is underway~\cite{CDS}.

Remarquably, these predictions only depend on the grossest features of the bubble collisions. Gravitational radiation is insensitive to the internal structure of colliding bubbles, in other words, to the small scale configuration of the scalar field in the colliding region~\cite{Kosowsky:1992rz}.  As confirmed by numerical simulations~\cite{Kosowsky:1991ua}, the {\it enveloppe approximation} works very well. It 
 consists in neglecting the dynamics of the collision (overlapping) region. Kinetic energy is supposed to be concentrated in the uncollided (but spherically asymmetric) bubble walls. 
Once bubble walls collide, they stir up the plasma at a scale comparable with their radii at the collision time, leading to turbulence which also induces gravitational emission.

\subsection{Key parameters characterizing the GW spectrum}

A crucial parameter for the calculation of the gravitational wave spectrum  is the rate of variation of the bubble nucleation rate, called $\beta$. This quantity fixes the characteristic scale in the problem, 
the size of bubbles at the time of the collision, and therefore the characteristic frequency $f_*$. 
The duration of the phase transition is given by 
$\beta^{-1}$ and the size of bubbles is typically   $R_b \sim v_b \beta^{-1}$ where $v_b$ is the velocity of the bubble wall. The initial size of the bubble at the time of nucleation (of the order of $T^{-1}$) is negligible compared to $\beta^{-1}$ which is of the order of the horizon size.

The second crucial parameter  characterizing the spectrum of gravitational waves is $\alpha=\epsilon / \rho_{rad}$, the ratio of the latent heat liberated at the phase transition ($\epsilon$=latent heat) to the energy density in the high energy phase, commonly being radiation energy density. $\epsilon$ is not necessarily vacuum energy, see for instance~\cite{RS}.
$\alpha$ and $\beta$ are evaluated at the nucleation temperature and determine entirely the GW spectrum.
They can be computed once we know the effective action for nucleating bubbles (``critical bubbles") which can be computed for any scalar potential describing the phase transition. Therefore, 
given a scalar potential at finite temperature, $V(\phi,T)$, this is enough to derive the predictions for the GW spectrum.

The rate of bubble nucleation is
\be
\Gamma(t)=A(t) e^{-S(t)}
\label{nucleationrate}
\ee
The prefactor $A(t)$ has units of energy to the fourth power,  $A(t)\sim {\cal M}^4$,
 where ${\cal M}\sim T$ is the typical energy scale of the transition and
 $S(t)\approx S_3/T$ is the euclidean action of the critical bubble. 
\be
S_3=\int 4 \pi r^2 dr \left[ \frac{1}{2}\left(\frac{d\phi_b}{dr}\right)^2 + V(\phi_b,T)\right]
\ee
is the free energy of a critical bubble.
$\phi_b$ is the bubble profile of the critical bubble obtained by solving for the ``bounce'':
\be
\label{bounce}
\frac{d^2\phi_b}{dr^2}+\frac{2}{r}\frac{d\phi_b}{dr}-\frac{\partial V}{\partial \phi_b}=0,  \ \ \  \mbox{with} \ \ \ \left. \frac{d\phi_b}{dr}\right|_{r=0}=0 \  \ \mbox{and} \ \ \left. \phi_b\right|_{r=\infty}=0
\ee
Most of the time variation of $\Gamma(t)$ is in $S(t)$, and $\beta$ is defined as:
\be
\beta \equiv -\left. \frac{d S}{d t}\right|_{t_*} \approx \left.\frac{1}{\Gamma}\frac{d\Gamma}{dt}\right|_{t_*}
\ee
where $t_*$ is the time when the transition completes. In a neighbourhood of $t_*$, $\Gamma(t)$ grows exponentially with time as $S(t)=S(t_*)-\beta (t-t_*) + ...$.
From the adiabaticity of the expansion of the universe, ${d T}/{dt}=-T H$, where $H$ is the expansion rate of the universe, and we obtain
\be
\frac{\beta}{H_*}=T_*\left. \frac{d S}{d T}\right|_{T_*}=T_* \left.\frac{d}{dT} \left(\frac{S_3}{T}\right)\right|_{T_*}
\ee
The temperature of the transition $T_*$ is defined as the temperature at which the probability for nucleating one bubble per horizon volume per horizon time approaches 1. This guarantees that bubbles percolate even if the universe is inflating. This translates into  
\be
\frac{\Gamma}{H^4} \sim O(1) \ \ \rightarrow S=-4 \ln \frac{{ T_*}}{m_{Pl}} 
\label{nucleationcondition}
\ee
$\beta/H $ is dimensionless and mainly depends on the shape of the potential at the time of nucleation. According to (\ref{nucleationcondition}), it depends only logarithmically on the energy scale. For a phase transition at the weak scale $\beta/H\sim 10^2$ and this justifies what we said after
 Eq.~(\ref{characteristic_frequency}) (where $f_*$ stands for $\beta$). 
 While the size of the bubble increases by orders of magnitude between nucleation and percolation
 (its initial radius $R\sim \phi/\sqrt{\Delta V}$ at the time of nucleation can be neglected; all what matters is the typical size at the end of the transition 
which is given by $R_b \sim v_b \Delta t \sim v_b \beta^{-1}$), $T_*$ (and thus $\alpha$ and $\beta/H$) is essentially unchanged between nucleation and percolation.

Note that the ratio $\beta/H $ also fixes the number of bubbles. The number density of bubbles is roughly $\beta^{-1}\Gamma(t)$ so that the number of bubbles in one horizon volume is $\beta^{-1}  \Gamma/H^3$, which, at the end of the transition is of order $H/\beta$. 

The latent heat is the sum of two contributions. The first one is 
the difference in free energies between the stable and metastable minima (which vanishes at $T_c$) while the second one comes from the entropy variation $\Delta {s}$ (which is non zero at $T_c$ in a first-order PT). This leads to the following formula for $\epsilon$:
\be
\epsilon=-\Delta V - T  \Delta { s}=\left(-\Delta V + T  \partial V/ \partial T \right)_{T_*}
\label{latentheat}
\ee

To conclude, for the analysis of gravity wave emission, the only two relevant quantities are the amount of latent heat injected into the plasma, Eq.~(\ref{latentheat}), and the nucleation rate, Eq.~(\ref{nucleationrate}).  So once  the
bubble action is computed (search for the bounce solution Eq.~(\ref{bounce})), everything is known.
 The fact that the phase transition is described or not by a  fundamental scalar field is irrelevant. 
 
In practise, instead of solving Eq.~(\ref{bounce}), one can use either the thin wall or thick wall limits to approximate the bounce solution. In the regime of large supercooling (which is the one of interest as far as large signals of gravitational waves are concerned), the thick wall approximation is adequate. In addition, if the temperature is decreasing and the ratio $T/T_c$ is getting low ($T_c$ is the critical temperature at which the free energies in the two phases are equal) one can use the 4D euclidean action $S_4$ to evaluate the nucleation rate rather than $S_3/T$. 

\subsection{Scaling expectations}

The energy density in gravitational waves coming from bubble collision can be estimated by naive dimensional analysis as follows.
The quadrupole formula for the power of gravitational emission is $ P_{GW}=\frac{G}{5}\langle ( {\stackrel{\ldots}{Q}} _{ij}^{TT})^2\rangle$ where $G$ is the Newton's constant and
$Q _{ij}^{TT}$ is the quadrupole moment of the source which is  $T _{ij}^{TT}$, the transverse traceless piece of the stress tensor.
We can write 
\be
 {\stackrel{\ldots}{Q}} _{ij}^{TT} \sim \frac{\mbox{mass of system in motion} \times (\mbox{size of system})^2}{(\mbox{time scale of system})^3}\sim \frac{\mbox{kinetic energy}}{\mbox{time scale of system}}
\ee
thus, $P_{GW} \sim G \dot{E}_{kin}^2$.
Let  $\kappa$ be the efficiency factor which quantifies the fraction of the vacuum energy  which goes into kinetic energy of bulk motions of the fluid (as opposed to heating):
\be
E_{kin}\sim \kappa \  \alpha \ \rho_{rad} \ (v_b \beta^{-1})^3
\ee
Using $G\sim H_*^2/ \rho_{tot*}$ and $\rho_{tot*}=(1+\alpha)\rho_{rad}$ we get ($d/dt$ leads to a $\beta^2$ factor)
$\rho_{GW*}=E_{GW}/(v_b^3 \beta^{-3})$ where $E_{GW}=P_{GW} \beta^{-1}$. Finally:
\be
\Omega_{GW*}=\frac{\rho_{GW*}}{\rho_{tot*}}\sim \left(\frac{H_*}{\beta}\right)^2\kappa^2\frac{\alpha^2}{(1+\alpha)^2}v_b^3
\ee
To get a large signal, we need ${\beta}/{H_*}$ to be small and  $\alpha$ to be large, in other words, the phase transition should last as long as possible and the latent heat should be maximized.
The scaling obtained is very close to what a more rigorous calculation gives for the redshifted value of the energy 
density evaluated at the peak frequency~\cite{Kamionkowski:1993fg}:
\be
\label{Omegacoll}
\Omega_{\mbox{\tiny coll}} \ h^2 (f_{\mbox{\tiny coll}}) \simeq 1.1 \times 10^{-6} \kappa^2 \left[\frac{H_*}{\beta}\right]^2 \left[\frac{\alpha}{1+\alpha}\right]^2\left[\frac{v_b^3}{0.24+v_b^3}\right]\left[\frac{100}{g_*}\right]^{1/3}
\ee
while the peak frequency is~\cite{Kamionkowski:1993fg}
\be
\label{fcoll}
f_{\mbox{\tiny coll}} \simeq 5.2 \times 10^{-3} \mbox{mHz} \left[\frac{\beta}{H_*}\right]\left[\frac{T_*}{100 \mbox{GeV}}\right]\left[\frac{g_*}{100}\right]^{1/6}
\ee
The scaling is slightly  different in the case of GW from turbulence in the plasma (the analysis 
of turbulent motions of~\cite{Kosowsky:2001xp} was generalized in~\cite{Dolgov:2002ra, Nicolis:2003tg}):
\be
\label{Omegaturb}
\Omega_{\mbox{\tiny turb}} \ h^2 (f_{\mbox{\tiny turb}}) \simeq 1.4 \times 10^{-4} u_s^5v_b^2 \left[\frac{H_*}{\beta}\right]^2 \left[\frac{100}{g_*}\right]^{1/3}
\ee
with~\cite{Dolgov:2002ra, Nicolis:2003tg}
\be
\label{fturb}
f_{\mbox{\tiny turb}} \simeq 3.4 \times 10^{-3} \mbox{mHz} \frac{u_s}{v_b}\left[\frac{\beta}{H_*}\right]\left[\frac{T_*}{100 \mbox{GeV}}\right]\left[\frac{g_*}{100}\right]^{1/6}
\ee
The turbulent fluid velocities $u_s$  are smaller than the bubble expansion velocities, unless turbulence is extremely strong. As a result, according to (\ref{fcoll}) and (\ref{fturb}), the peak  for the turbulence spectrum will be shifted to lower frequencies. Note that the peak frequency of the collision signal does not depend on $\alpha$. Because $f_{\mbox{\tiny turb}} \approx f_{\mbox{\tiny coll}}   \times (u_s/v_b)$ and $(u_s/v_b)$ is an increasing function of $\alpha$, as $\alpha$ increases, the collision peak gets hidden by the high-frequency tail of the turbulence signal\footnote{Although the relation $f_{\mbox{\tiny turb}} \approx f_{\mbox{\tiny coll}}   \times (u_s/v_b)$ is likely to be revised in Ref.~\cite{CDS}.} . 

The signal from turbulence is more promising than the signal from bubble collision due to the different scaling with $\alpha$. Indeed, the scaling with $\alpha$ of the velocities and the efficiency  factor are :
\be
v_b(\alpha)=\frac{1/\sqrt{3} +\sqrt{\alpha^2+2\alpha/3}}{1+\alpha} \  \mbox{~\cite{Steinhardt:1981ct}}  \ \ \  ,  \ \ \ u_s(\alpha)\simeq \sqrt{\frac{\kappa \alpha}{\frac{4}{3}+\kappa \alpha}}  \mbox{~\cite{Nicolis:2003tg}}
\ee
\be
\kappa(\alpha)\simeq \frac{1}{1+0.715 \alpha}\left[0.715 \alpha +\frac{4}{27}\sqrt{\frac{3\alpha}{2}}\right] \mbox{~\cite{Kamionkowski:1993fg}}
\ee

The formulae (\ref{Omegacoll})  and (\ref{Omegaturb}) for $\Omega h^2$ show that the amplitude of the signal does not depend on the energy scale of the transition but only on the dimensionless parameters ${\alpha}$ and ${H_* / \beta}$, in other words on  the shape of the scalar potential at the time of nucleation. Very roughly, taking $\beta/H_* \sim S_3(T_*)/T_*\sim {\cal O}(100)$, it is clear that we need $\alpha \sim {\cal O}(1)$ if we want $\Omega h^2 \gsim 10^{-10}$ (to see something at LISA). 
Our experience with possible values of $\alpha$ and $\beta/H_*$ is based on the studies of potentials describing the electroweak phase transition. The values of $\alpha$ and $\beta/ H_*$ are related to the ratio $\phi (T_*)/T_*$, another quantity characterizing the {\it strength} of the phase transition, where $\phi$ is the vacuum expectation value of the Higgs. 
For instance, in the Minimal Supersymmetric Standard Model (MSSM), $\alpha$ is typically smaller than 0.1 while $\beta/H_*$ is larger than 1000~\cite{Apreda:2001us}. On the other hand, in the NMSSM, the authors of~\cite{Apreda:2001us} found values of $\alpha \sim {\cal O}(1)$.
Values of $\alpha$ larger than 1 correspond to a phase transition which is so strong that it is at the borderline of not being able to take place because the free energy of a critical bubble is too large: either the barrier separating the two minima is too large (the barrier even exists at zero temperature) or the distance in field space between the two minima of the potential is too big. The same applies if $\beta/H_*$ is smaller than ${\cal O}(100)$.
 This situation is encountered not only in the NMSSM but also in effective theories with large negative quartic couplings~\cite{Grojean:2004xa}, as will be presented in~\cite{Delaunay}. Note that for a given scalar potential, $\alpha$ and $\beta/H_*$ are actually correlated as a large value of $\alpha$ will be associated with a small value of $\beta/H_*$. Indeed, $\alpha$ is proportional to the latent heat, which grows as $\Delta V$, the depth of the potential at the minimum, is getting bigger. On the other hand, $S_3/T$ (and thus $\beta/H_*$) typically scales like $1/\sqrt{\Delta V}$.

 Let us make a few comments  concerning the situation where $\alpha \sim 1$. In usual phase
  transitions,
  this means that the vacuum energy is of the same order as the radiation energy density, therefore inflation starts before bubbles percolate.  However, if Eq.~(\ref{nucleationcondition}) is satisfied, this guarantees that the phase transition can still complete. The number of e-foldings of inflation is 
  $\ln (V^{1/4}/T_*)$, which is  typically less than 1. However, if the transition is very slow, i.e. $\beta/H \lsim 1$, one should take formulae (\ref{Omegacoll}) and (\ref{Omegaturb}) with caution as they are derived neglecting the expansion of the universe, which is not a good approximation if $\beta/H<1$.
  
\begin{figure}[!htb]
\begin{center}
\includegraphics[height=7.cm,width=8.cm]{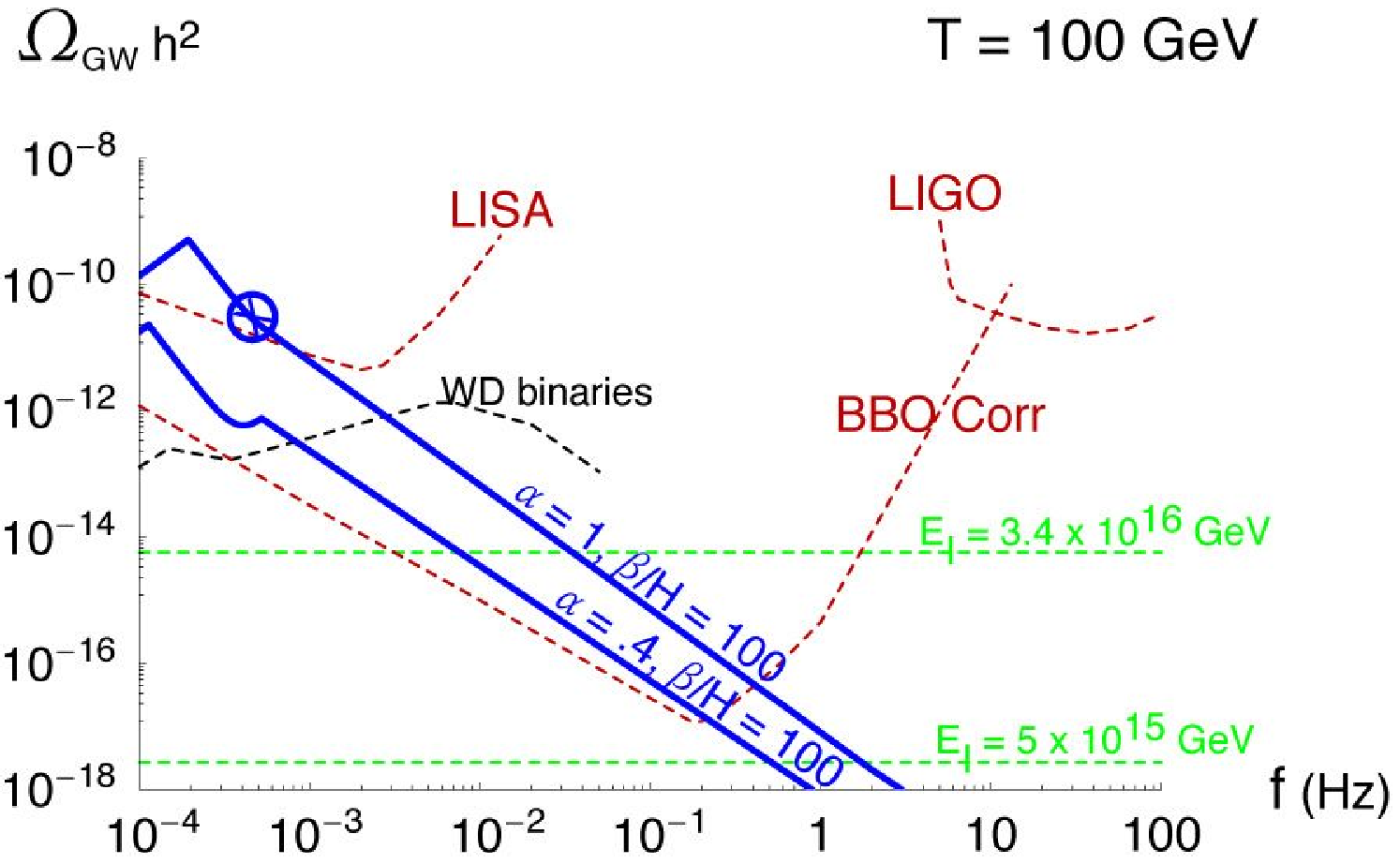}
\includegraphics[height=7.cm,width=8.cm]{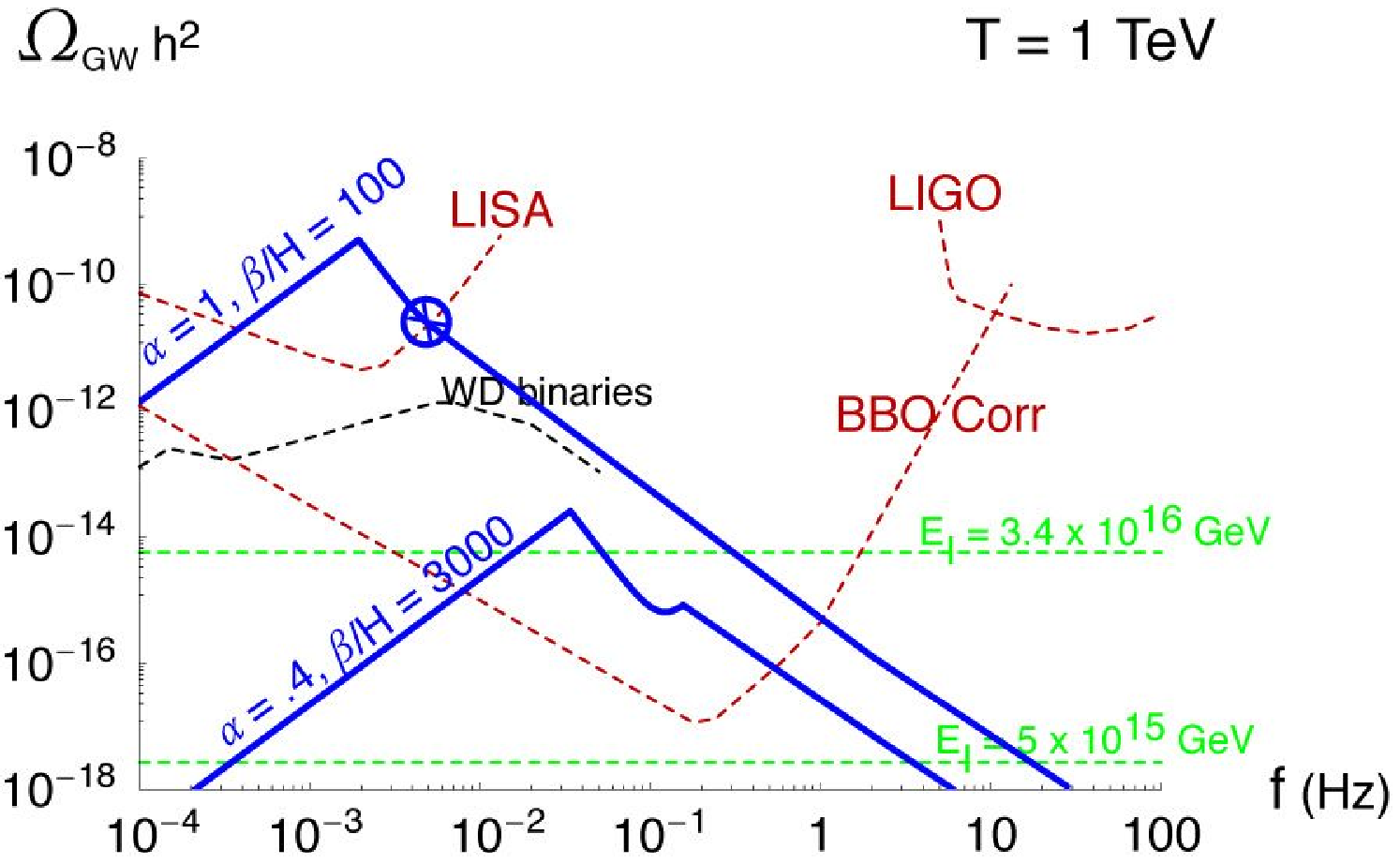}
\includegraphics[height=7.cm,width=8.cm]{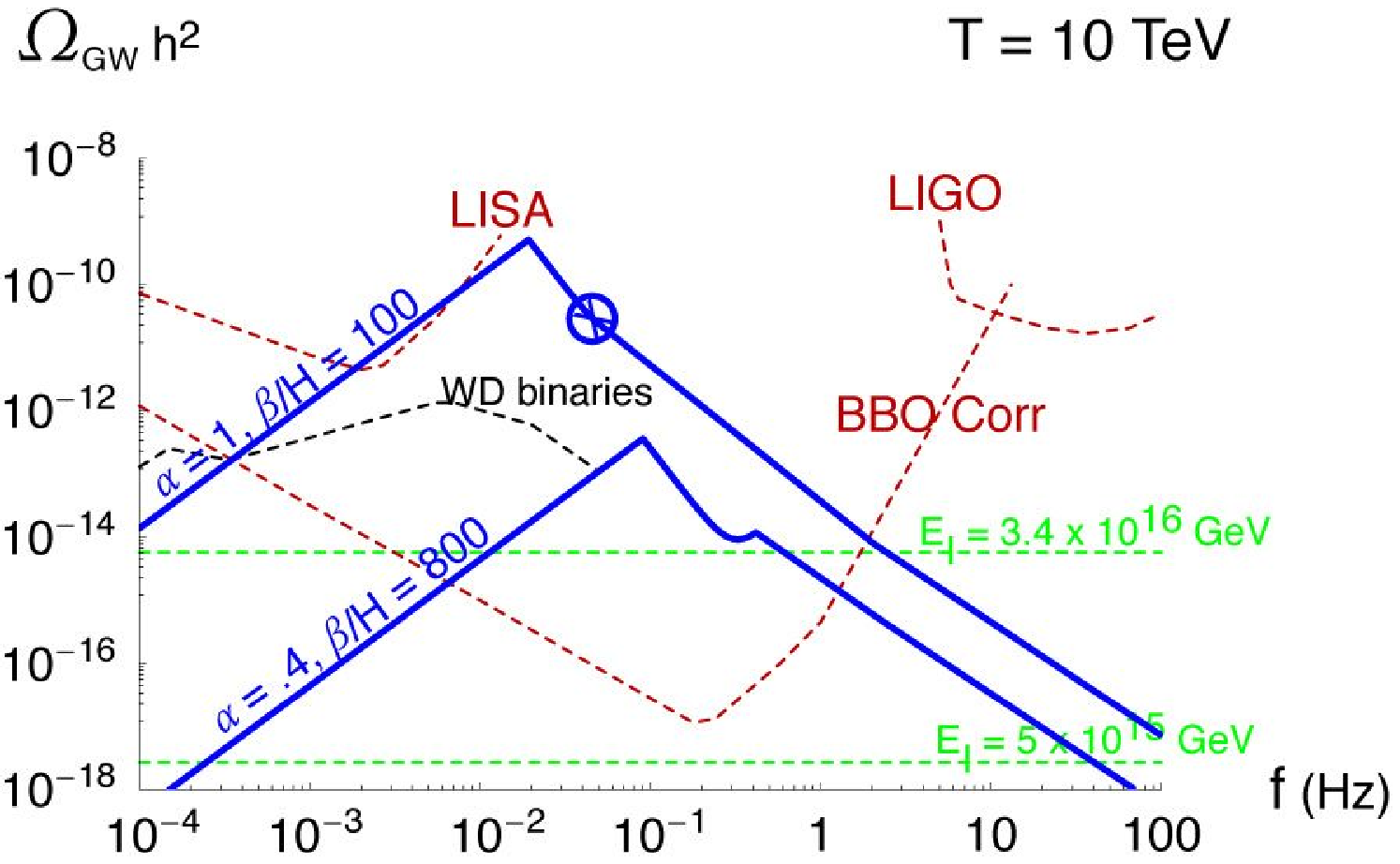}
\includegraphics[height=7.cm,width=8.cm]{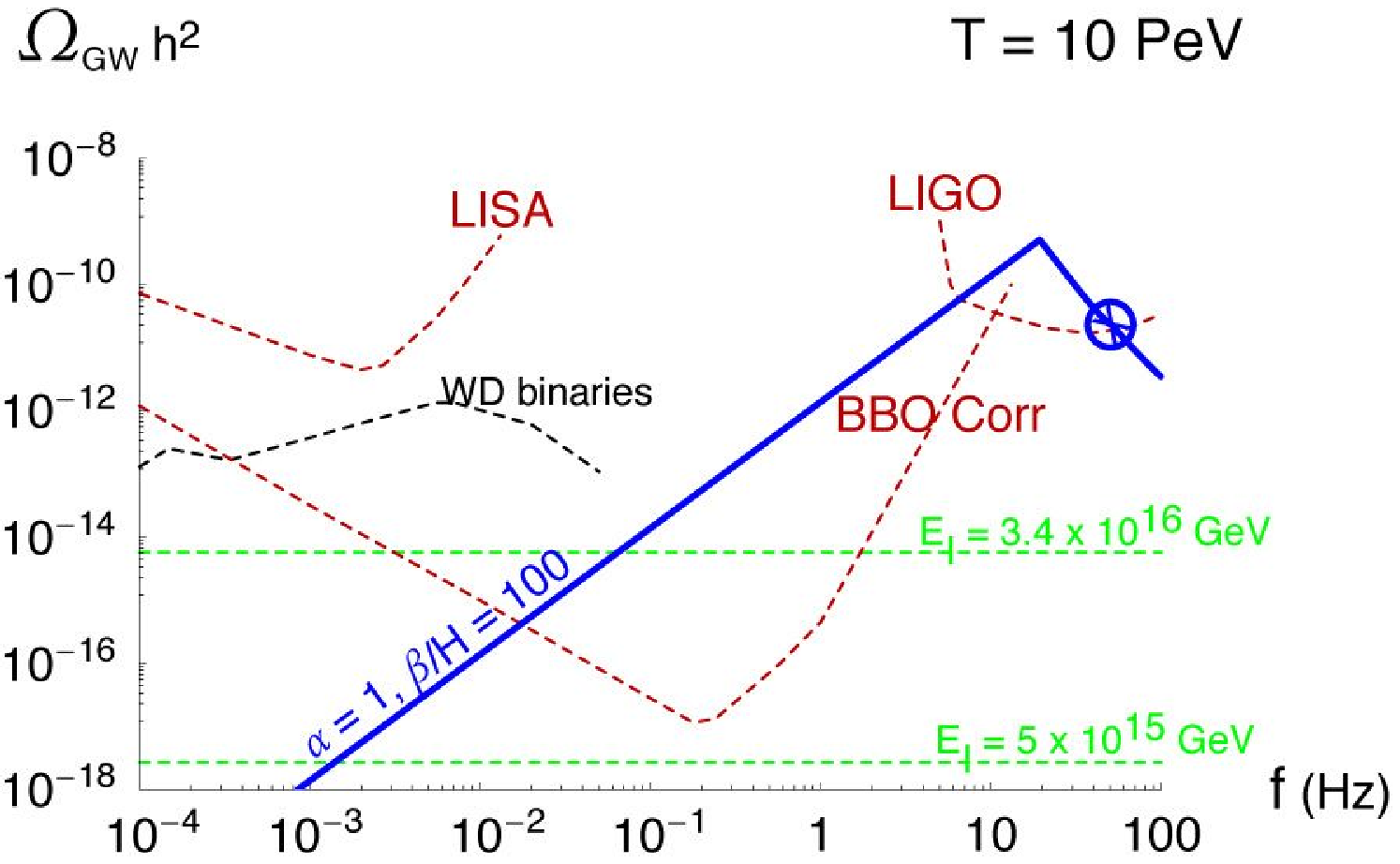}	
\caption{Spectrum of gravitational waves expected from a first order phase transition (solid blue line) for four temperatures and for some choices of ($\alpha, \beta/H$) values. The dashed red lines are the (approximate) predicted sensitivities of LISA, BBO, LIGO-III. The horizontal dashed green lines are the gravitational spectra expected from inflation, for two scales of inflation, for comparison. The black dashed curve is the estimate for the irreducible foreground due to white dwarf binaries (from~\cite{Farmer:2003pa}). At large $\alpha$, only the peak from turbulence can be seen as well as a change of slope (shown as a circled cross) corresponding to the high frequency tail of the bubble collision spectrum. For low $\alpha$, it is possible to see the collision peak as well.}
\label{fig:100GeV}
\end{center}
\end{figure}

  Fig.~\ref{fig:100GeV} shows some GW spectra illustrating the predictions for various temperatures with representative values of $\alpha = 0.4, 1$ and $\beta/ H_*=100, 800, 3000$. Those plots also exhibit
two examples of signals from inflation (and taken from~\cite{Smith:2005mm}) for comparison.
The scale of inflation is constrained by the CMB to be $E_{I}\lsim 3.4 \times 10^{16 }$~GeV~\cite{Smith:2005mm}. This fixes the largest signal we could expect from inflation as $\Omega_{\mbox{\tiny GW}} h^2\propto E_I^4$. We also include the signal corresponding to $E_{I} = 5 \times 10^{15 }$~GeV which could be observed at BBO.
  
\section{Scanning the ($\alpha$, $\beta/H_*$) plane}
In our analysis below, we will use the formulae of the previous section as well as the fact that the spectrum $\Omega_{\textrm{\tiny coll}} h^2 $ is expected to increase as $f^{2.8}$ while at high frequencies it drops off as $f^{-1.8}$ and that $\Omega_{\textrm{\tiny turb}}  h^2$ increases as $f^{2}$ while at high frequencies it drops off as $f^{-7/2}$. This is already all summarized in the letter~\cite{Nicolis:2003tg}. However,  Ref.~\cite{Nicolis:2003tg} focuses on the detectability at LISA of GW from a $T=100$~GeV phase transition. We are now using this formalism to look in more details at the detectability of GW coming from any other 1st order phase transitions at future interferometers.
We repeat that  we are working at the level  of an-order-of-magnitude estimate.
Magnetohydrodynamical effects could make the slope of the turbulence high frequency tail smaller~\cite{Nicolis:2003tg} and in any case, for a more precise analysis, the calculation of the power spectrum should be revisited first.
We compare the GW spectra resulting from PT occurring at temperatures in the range $[100~\mbox{GeV}, 100~\mbox{PeV}]$ with the sensitivities of LISA, BBO and LIGO correlated third generation (and taken from
\cite{Buonanno:2004tp}). 
The BBO sensitivity is approximate and may change in the final design. We are actually using the sensitivity of BBO Corr, its correlated extension, which correlates two detectors, namely two LISA-like constellations (each LISA-like constellation orbits around the Sun at 1 AU and consists in three spacecrafts in a triangular
configuration) that will allow to do correlations to measure the
stochastic background. As discussed in Section~\ref{astro_back}, we take into account as well the irreducible background due to extragalactic white dwarf (WD) binaries.

 \begin{figure}[!htb]
\begin{center}
\includegraphics[height=7.cm]{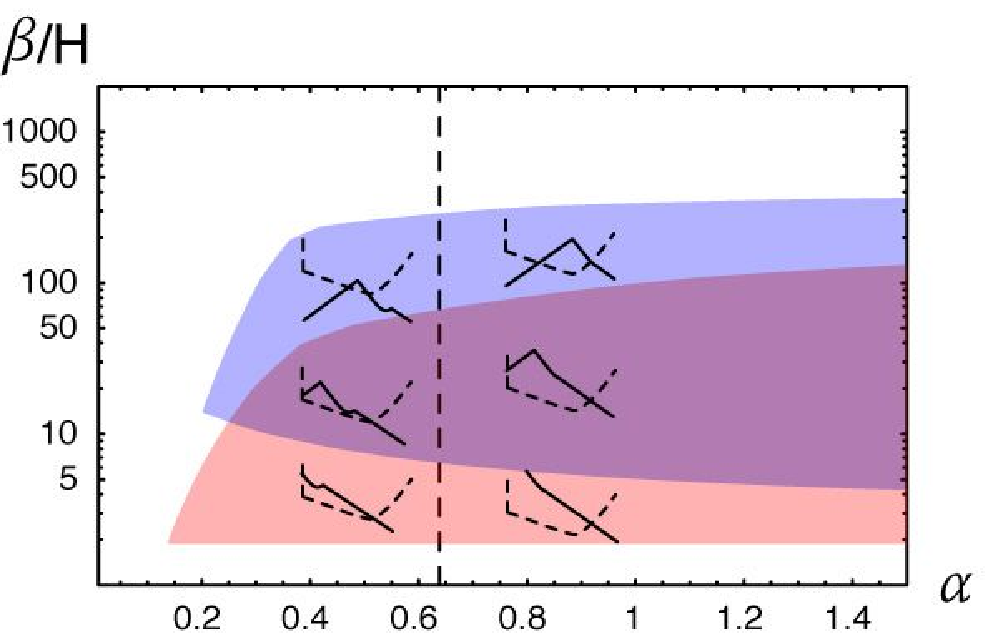}
\caption{Different configurations of the signal versus the instrument sensitivity to show the qualitative dependence on parameters. The upper blue region is where the turbulence peak is observable while the lower red one is the region where either the collision peak or the point of slope change is  visible.
Precise locations of these different regions depend on the experiment and the temperature of the transition as illustrated in Figs.~\ref{fig:LISAcontours}, \ref{fig:BBO} and~\ref{fig:LIGOcontours}.}
\label{fig:regions}
\end{center}
\end{figure}

For each temperature, we are making a full scan of the ($\alpha$, $\beta/H_*$) parameter space and determine the regions where at least one of the peaks is observable.
According to Eq.~(\ref{Omegacoll}, \ref{fcoll},\ref{Omegaturb},\ref{fturb}), various situations can arise: 

\begin{itemize}
\item For relatively low $\alpha$, the turbulence and collision peaks are well separated and can be observed. This is the ideal situation as the observability of these two peaks would be a smoking gun for the phase transition origin of these GW. The ratio of the two peak frequencies is a predicted function of $\alpha$. In somes cases, the turbulence peak is at too low frequency to be observed by LISA or BBO but the minimum separating the two peaks is visible.

\item At larger $\alpha$ ($\gsim$ 0.64), the collision peak is hidden by the high frequency tail of the turbulence peak. However, there is a characteristic change of slope in the high frequency tail. Depending on the temperature of the transition, this change of slope can be observed or not.
\end{itemize}

\begin{figure}[!htb]
\begin{center}
\includegraphics[height=4.9cm,width=8.cm]{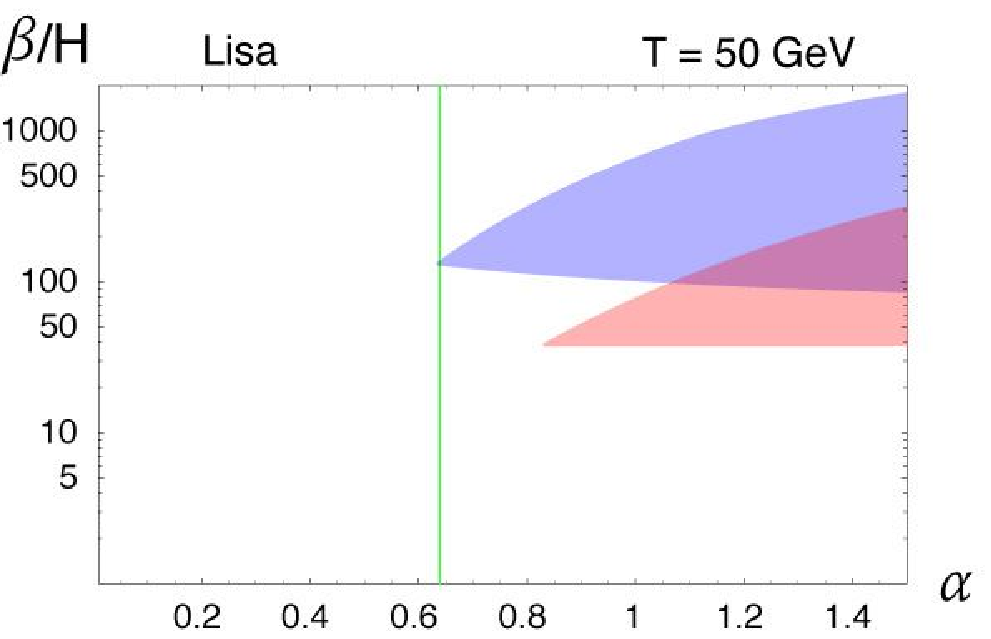}
\includegraphics[height=4.9cm,width=8.cm]{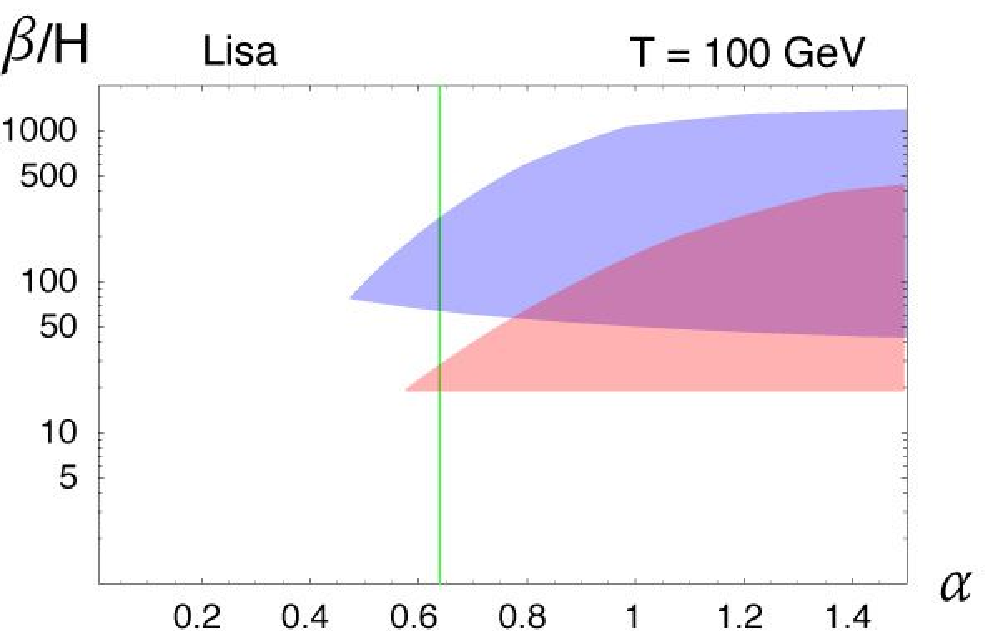}
\includegraphics[height=4.9cm,width=8.cm]{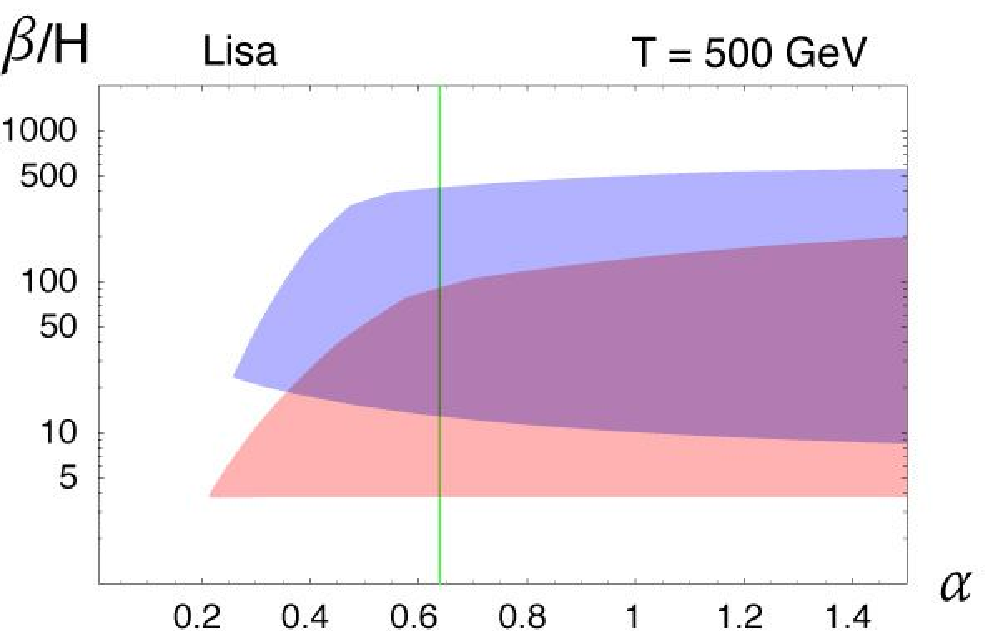}
\includegraphics[height=4.9cm,width=8.cm]{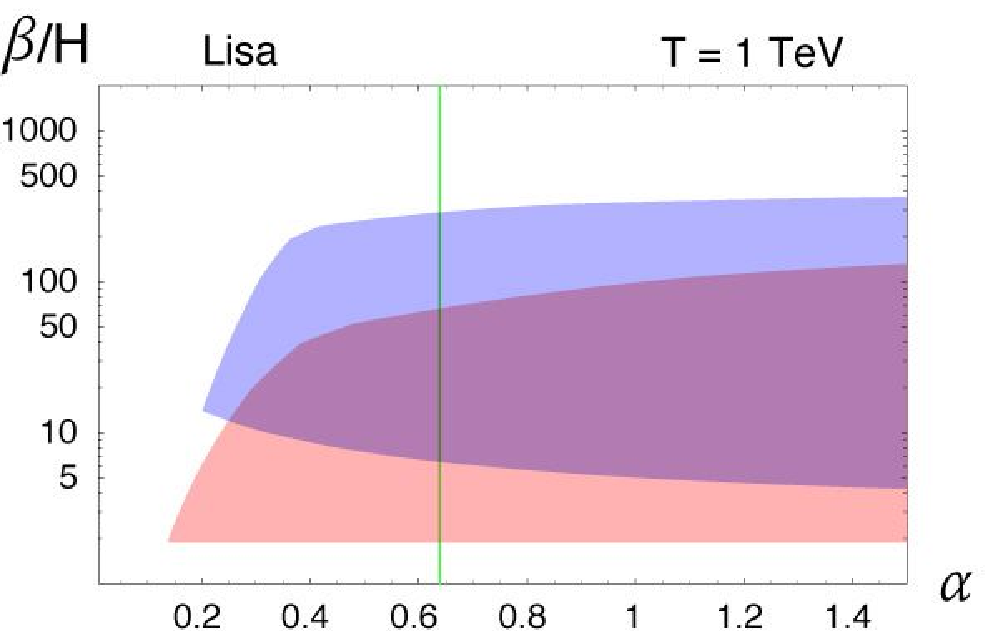}
\includegraphics[height=4.9cm,width=8.cm]{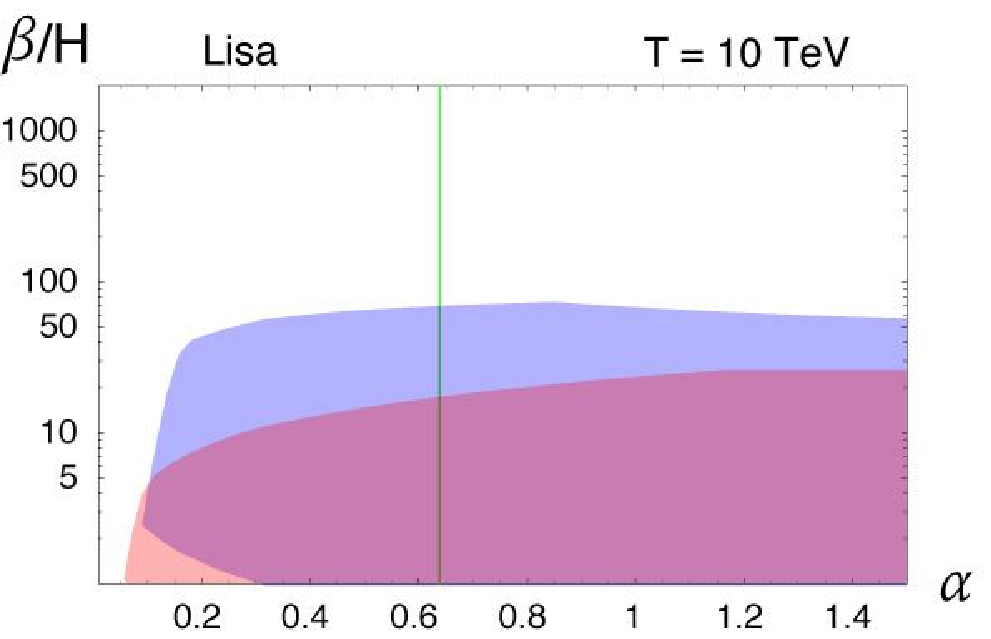}
\includegraphics[height=4.9cm,width=8.cm]{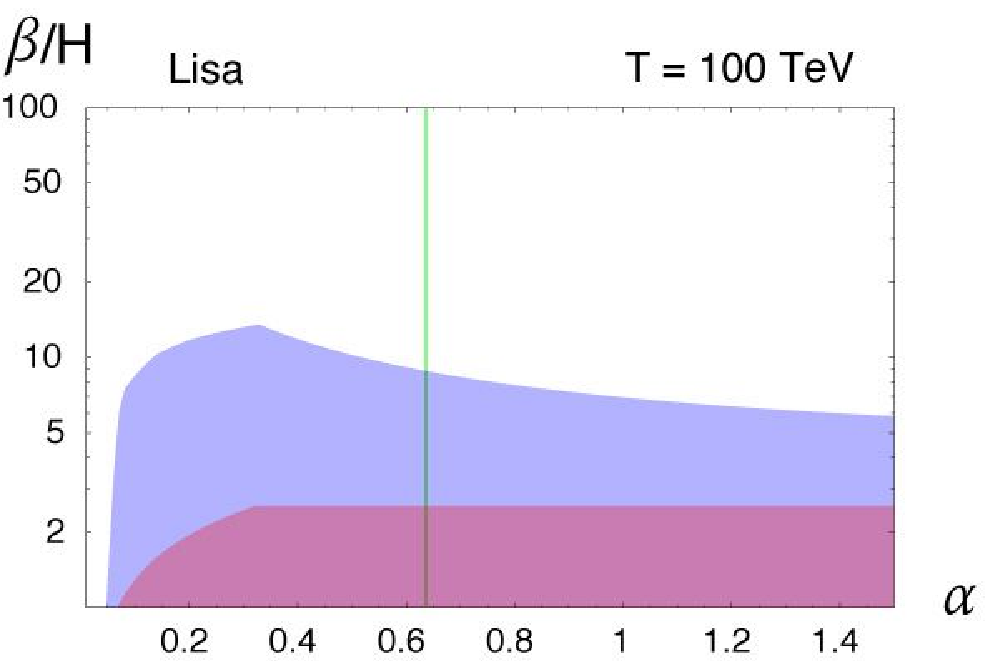}
\includegraphics[height=4.9cm,width=8.cm]{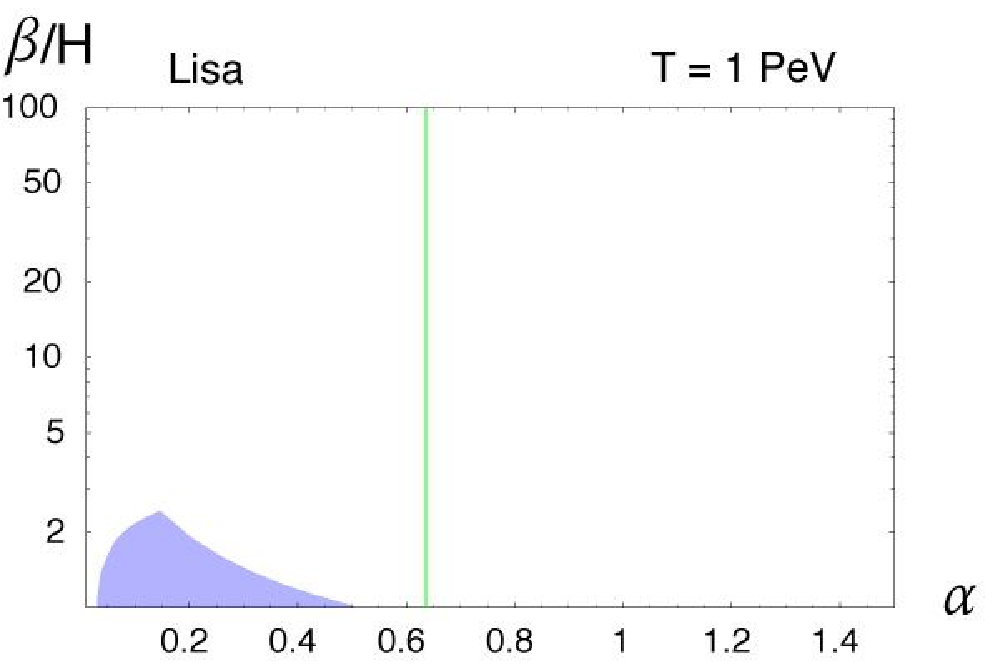}
\caption{Contours delimiting the region in the ($\alpha$, $\beta/H$) plane for which there is an observable peak at LISA. The upper blue region is for the turbulence peak while the lower red one is the region where either the collision peak or the point of slope change is  visible. Left of the vertical green line, the collision peak is visible.}
\label{fig:LISAcontours}
\end{center}
\end{figure}

\begin{figure}[!htb]
\begin{center}
\includegraphics[height=4.9cm,width=8.cm]{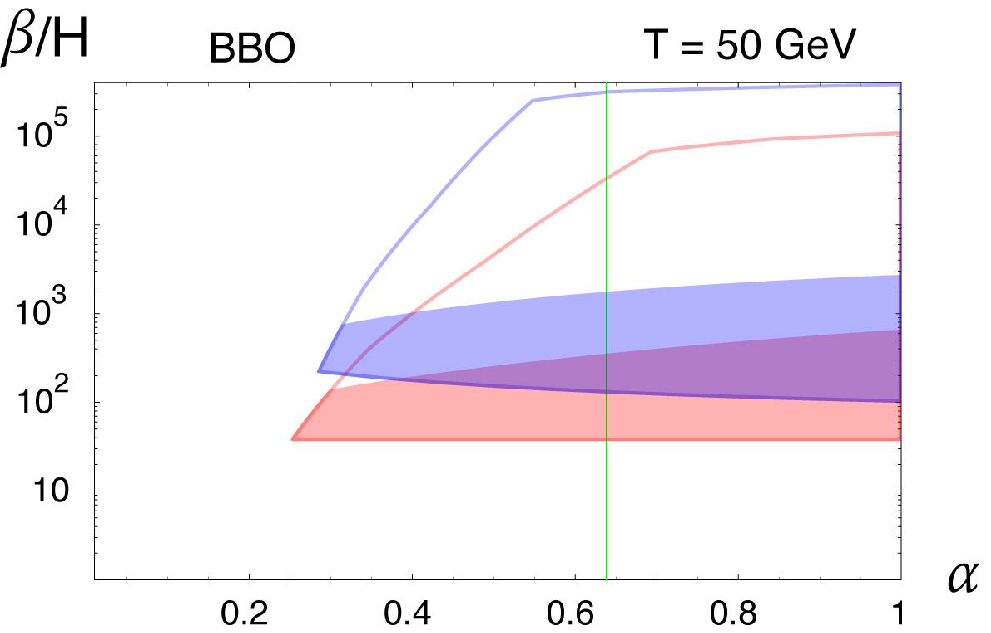}
\includegraphics[height=4.9cm,width=8.cm]{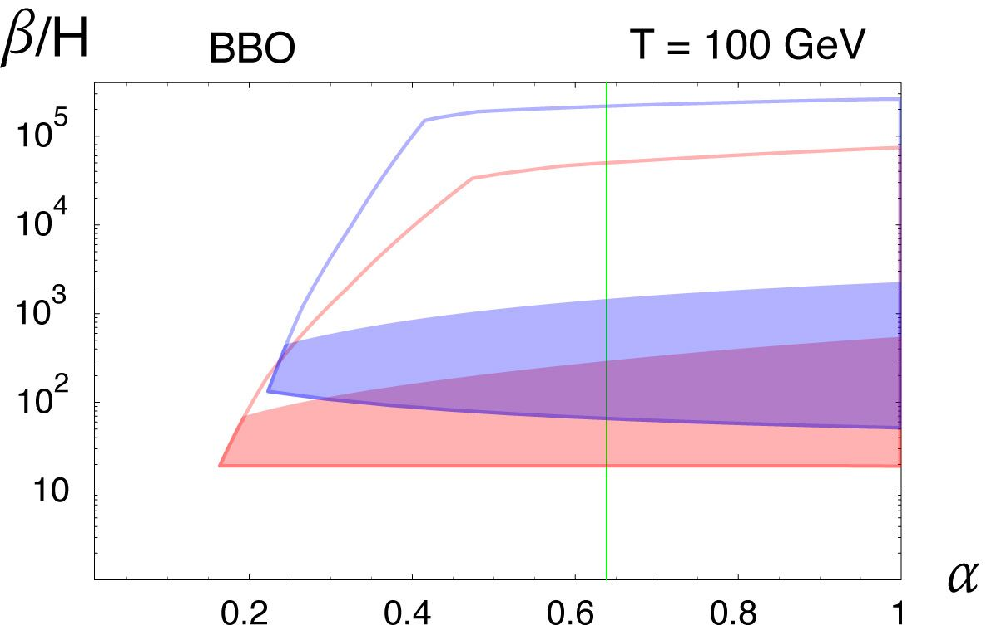}
\includegraphics[height=4.9cm,width=8.cm]{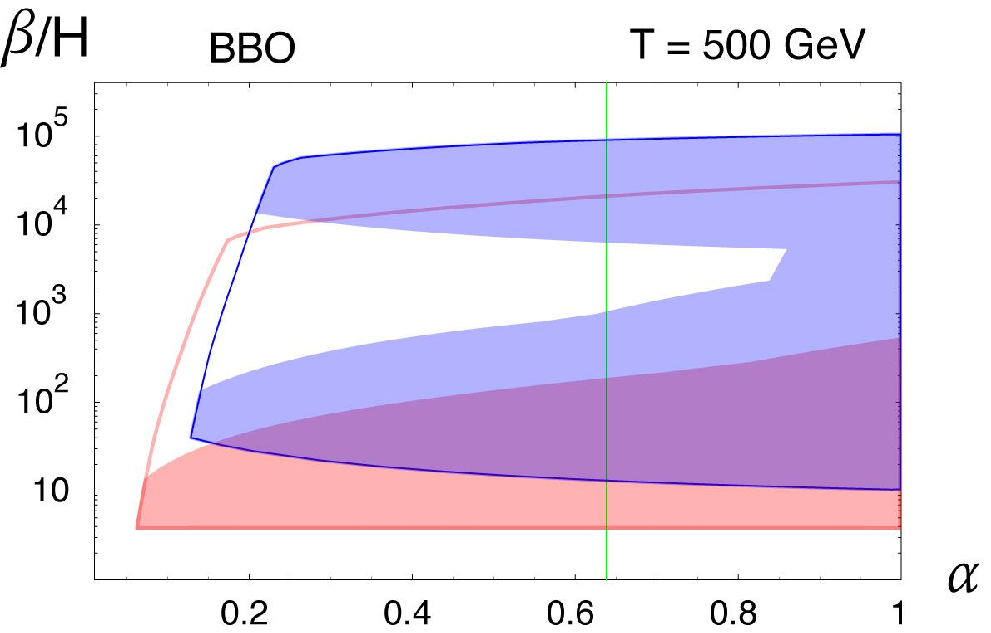}
\includegraphics[height=4.9cm,width=8.cm]{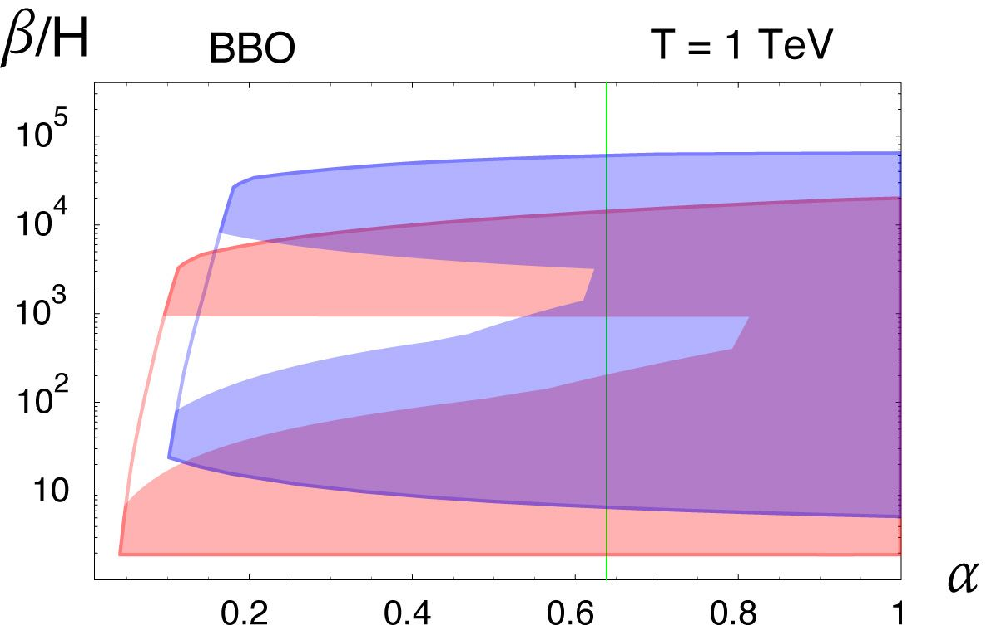}
\includegraphics[height=4.9cm,width=8.cm]{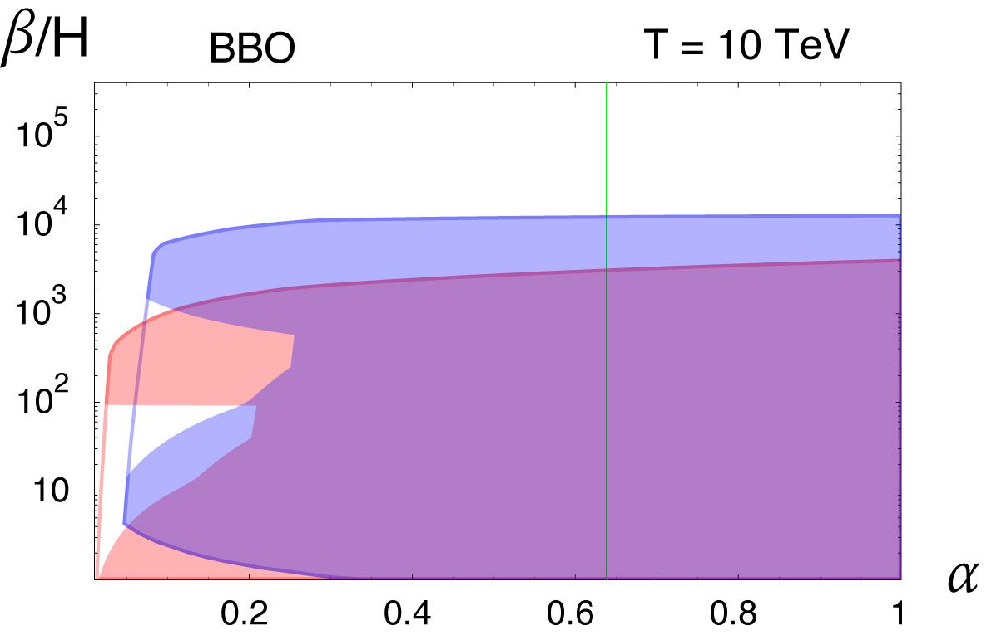}
\includegraphics[height=4.9cm,width=8.cm]{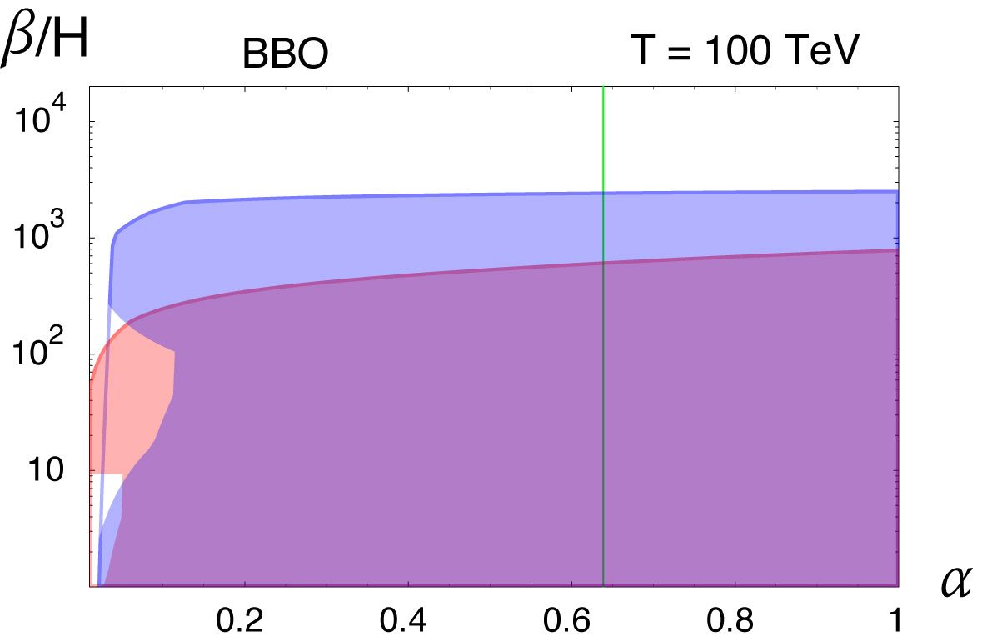}
\includegraphics[height=4.9cm,width=5.33cm]{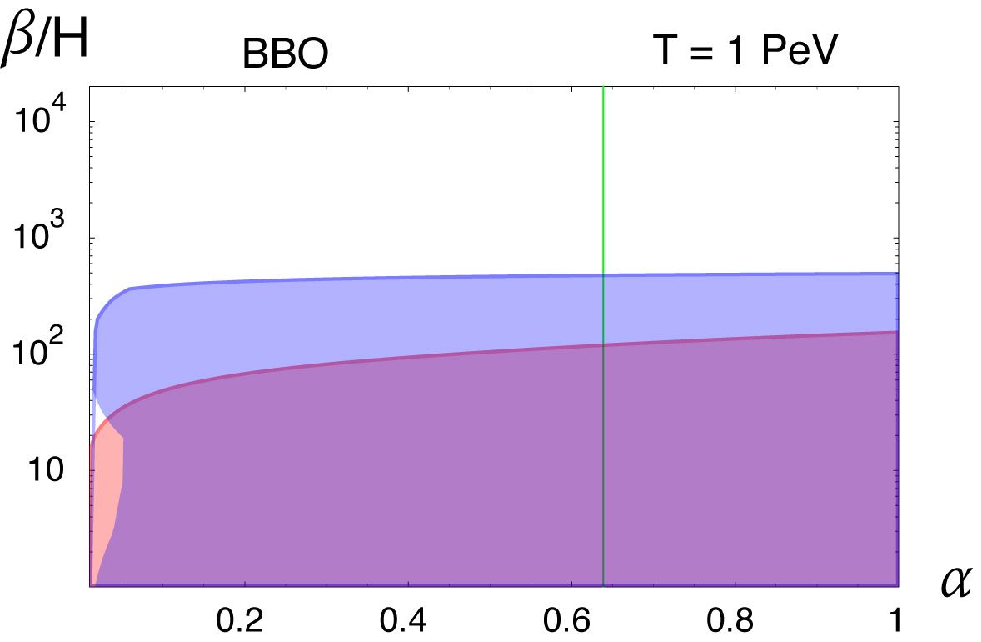}
\includegraphics[height=4.9cm,width=5.33cm]{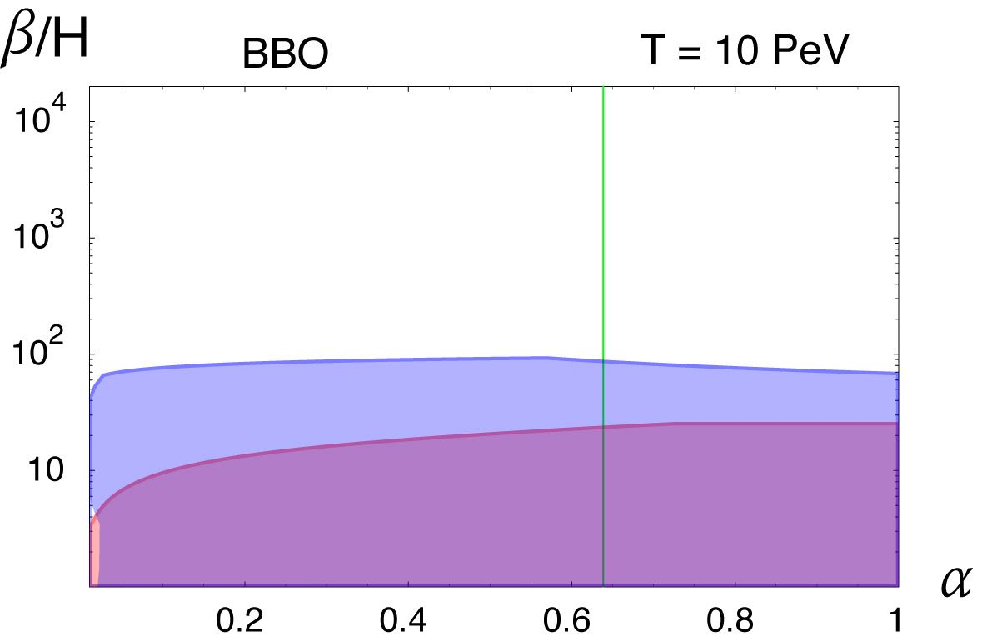}
\includegraphics[height=4.9cm,width=5.33cm]{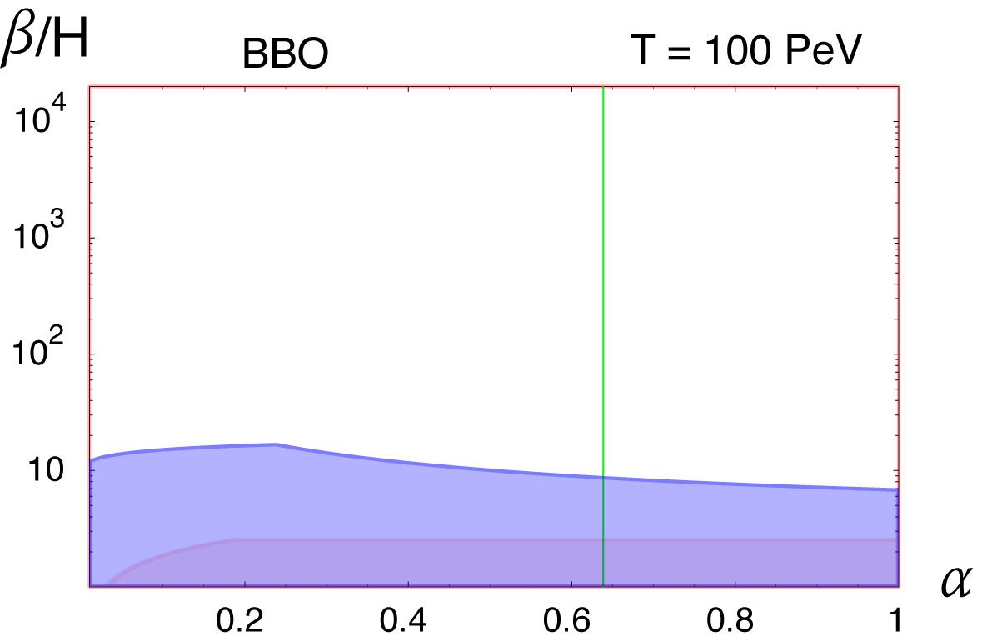}
\caption{Same as Fig.~\ref{fig:LISAcontours} but for BBO. The effect of including the constraint from the irreducible WD foreground is displayed and limits the observable regions from the uncolored ones to the ones in plain colors. As the temperature increases, the peaks are shifted to higher frequencies, thus the effect of the WD foreground becomes less significant.}
\label{fig:BBO}
\end{center}
\end{figure}

\begin{figure}[!htb]
\begin{center}
\includegraphics[height=5.cm,width=8.1cm]{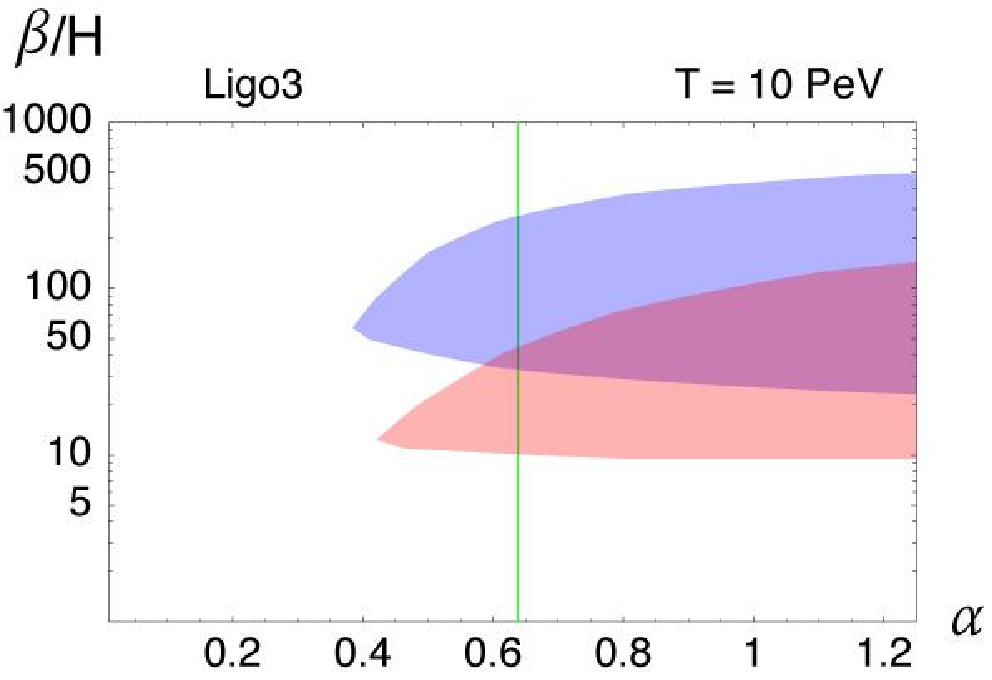}
\includegraphics[height=5.cm,width=8.1cm]{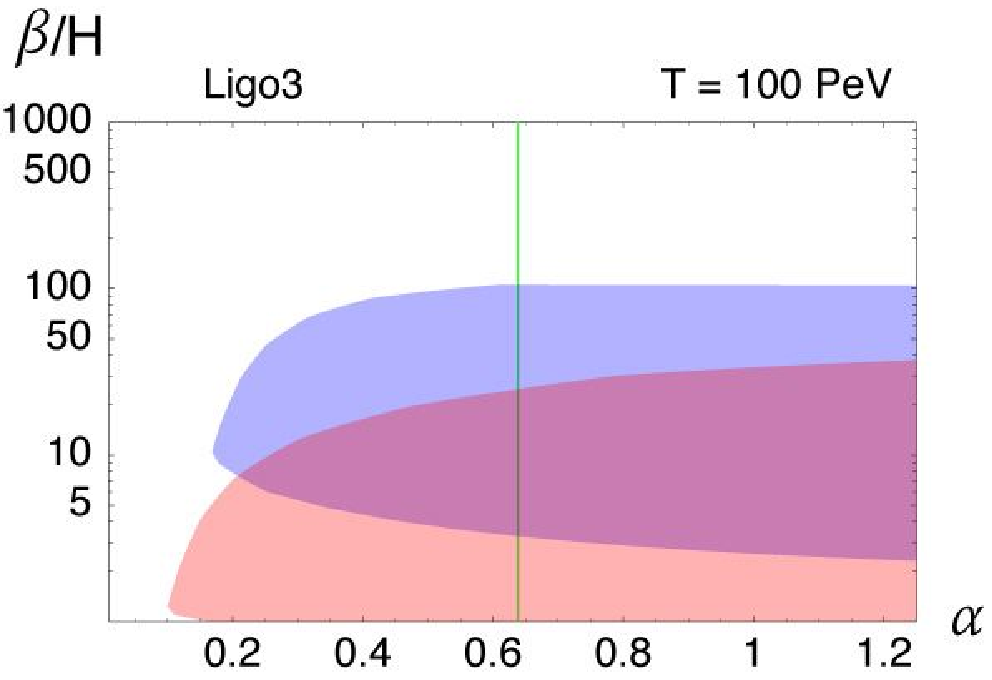}
\caption{Same as Fig.~\ref{fig:LISAcontours} but for LIGO-III.}
\label{fig:LIGOcontours}
\end{center}
\end{figure}

Our contour plots show the region where the turbulence peak is observable and the region where either the collision peak or the slope change is visible, at LISA (Fig.~\ref{fig:LISAcontours}), BBO (Fig.~\ref{fig:BBO}) and LIGO (Fig.~\ref{fig:LIGOcontours}). The vertical line separates the low $\alpha$ region where the two peaks are well separated  from the large $\alpha$ region where only the change of slope is visible. The lower horizontal bound is due to the fact that we cut the sensitivity of both LISA and BBO at $10^{-4}$ Hz. Because of this frequency limit in the sensitivity, we cannot probe phase transitions below a $\sim$ GeV and thus the QCD phase transition.  In the BBO plots, we show the very important effect of the WD foreground on the detectability at BBO.
 
 Note that we could have also made contours corresponding to cases where none of the peaks are observable but the high or low frequency tails can still be detected. This will clearly enlarge the detectability region and this is work in progress.

\begin{figure}[!htb]
\begin{center}
\includegraphics[width=12cm]{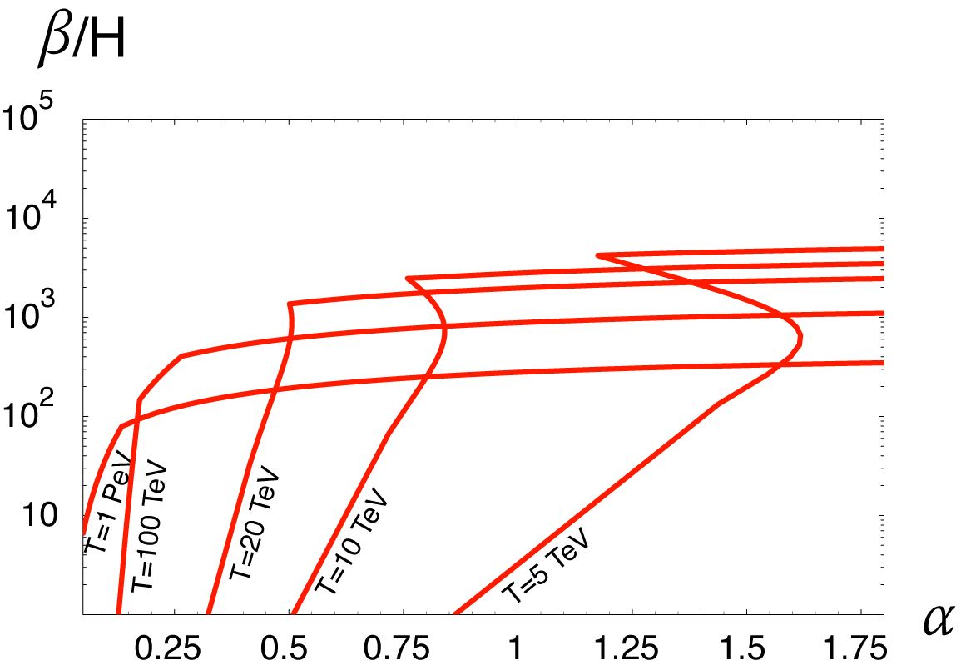}
\caption{Below each line (each one associated with the temperature of the phase transition), the gravitational wave signal at BBO from 1st order phase transitions entirely masks the signal expected from inflation. This plot strongly depends on the scale of inflation, which was chosen here to be $E_I= 3.4 \times 10^{16}$~GeV.}
\label{screen}
\end{center}
\end{figure}

\subsection{ T=100 GeV}

LISA will be able to detect the peak of GW from a 100~GeV first order PT only if it is extremely strong ($\alpha \gsim 0.5$ or $\beta/H \lsim 1000$).  
The Higgs potential of the MSSM does not satisfy this requirement but it can in the NMSSN~\cite{Apreda:2001us}. Higgs potentials with negative quartic couplings can also trigger strong EWPT as was shown in~\cite{Grojean:2004xa}. The corresponding prospects for GW detection will be presented in details elsewhere~\cite{Delaunay}. There are also exciting large signals expected from the high temperature behaviour of a warped extra dimension~\cite{RS}.
If $T$ is $\sim 500$~GeV rather than 100~GeV, the GW peak will coincide with LISA's best sensitivity frequency and a larger region will be detectable.
The prospects for detection of GW from the 100~GeV EWPT are very good at BBO. At these low frequencies, the WD foreground is lower.  For instance, the turbulence peak for $\alpha \gsim 0.3$, 
${\beta}/{H}\sim 200$, is above the WD foreground and can be seen by BBO. 
Note also that at $T=500$~GeV, values of $\beta/H$ as large as $10^4$--$10^5$ can  be probed.

\subsection{T=1 TeV}

It is quite exciting that a 1 TeV PT with $\alpha \gsim 0.4$, ${\beta}/{H} \sim 200$ can be seen by LISA. 
Recent radical proposals to address the hierarchy problem predict rich new phenomena at the TeV scale. For instance, new dimensions at a TeV could give rise to observable signals at LISA~\cite{RS}. 
There was also a recent study of the high temperature behavious of Little Higgs Theories where it was shown that EW symmetry was restored precisely at a temperature of order $\sim$ 1 TeV~\cite{Espinosa:2004pn}. This transition appeared to be first order. A more detailed analysis would be required to determine whether this PT could be strong enough to lead to an observable spectrum of GW at LISA. Unfortunately, this takes place in the regime where the effective theory ceases to be under control. This is  nevertheless an interesting prospect.

If ${\beta}/{H} \gsim 100$, the turbulence peak is above the  WD foreground for $\alpha \gsim 0.2$,  and thus can be seen by BBO. In addition, BBO can see the high frequency tail of the collision peak which covers good part of, if not entirely,  the inflation signal.

\subsection{T=10 TeV}
At these temperatures, the peak cannot be probed by LISA, unless ${\beta}/{H} < 10^2$. On the other hand, LISA can still probe the low frequency tail of these spectra and is therefore a compelling tool to probe scales that LHC will not be able to reach.
As the temperature increases, the peaks are shifted to higher frequencies, thus the effect of the WD foreground becomes less significant and quite weak first order phase transitions can be probed.
And the high frequency tail of the collision peak can entirely screen the inflation signal, depending on the scale of inflation (see Fig.~\ref{screen}). 
If ${\beta}/{H}\sim 1000$, it is possible to see both the turbulence and the collision peaks for $0.1 \lsim \alpha \lsim 0.64$ and assuming that the inflationary scale is sufficiently low. For ${\beta}/{H} \sim 200$, the collision peak can be seen for $\alpha $ as low as $\sim $0.05 if  the inflationary signal is below the BBO sensitivity. 
 
\subsection{T=100 TeV}

If ${\beta}/{H}\sim 200$ and $\alpha \gsim 0.2$ the inflation signal is for sure entirely covered.  If the inflation scale is below $\sim 5 \times 10^{15}$~GeV,  $\alpha $ as low as 0.05 could be detected at BBO.
Two peaks can be seen if $0.1 \lsim  \alpha \lsim 0.64$  and ${\beta}/{H}  \sim$ 200. At larger ${\beta}/{H}$ only the turbulence peak will be seen.

\subsection{T=$10^7$ GeV}

This is a particularly interesting case as the same signal could be observed by both BBO and Ligo-III. Specifically, a phase transition with $\alpha=0.8$ and ${\beta}/{H}  \sim$ 200 would give a turbulence peak observable by Ligo-III while the low frequency tail would be observable by BBO. In this example, the inflation signal would be hidden except in a very narrow frequency range between 50 mHz and  80 mHz. \\

Interesting signatures end at this energy scale. Phase transitions at T$\gsim 10^8$~GeV cannot be probed by any of the planned interferometers.

\section{Conclusion}
We have shown that the GW background from early universe phase transitions may become relevant for a second generation detector such as the Big Bang Observatory (BBO) which is so far motivated to detect the GW background produced during inflation.
LISA, LIGO and BBO will be able to probe part of the history of the universe in the temperature range 100~GeV--$10^7$~GeV. The GW signal coming from particle physics phase transitions is directly related to the scalar potential describing the evolution of the order parameter. Observation or non-observation of GW will allow to put constraints on the parameters of these potentials.
The measurement of the GW spectrum (peak frequency and intensity) can discriminate among different models (once combined with experimental measurement at colliders, for instance, once knowing the higgs mass) and put constraints on the model parameters. For example, at LHC, we will be able to measure the Higgs mass but not the quartic or cubic self coupling of the Higgs.  Only a linear collider can provide this information, which timescale could be beyond LISA. LISA could start constraining model parameters before a linear collider. In addition, LISA is sensitive to the 10 TeV scale which is beyond the reach of the near future collider experiments.

The gravitational wave signal from phase transitions at around 10--100~TeV temperatures could entirely screen the signal from inflation, which detection is one of the main motivations for building BBO.

We emphasize that our quantitative analysis can only be indicative given the uncertainties both at the experimental and theoretical level. The sensitivities of LISA, LIGO-III and BBO will certainly change during the next years. On the theoretical side, we use the estimate of the GW power spectrum of~\cite{Kosowsky:1992rz,Kosowsky:1991ua,Kosowsky:1992vn,Kamionkowski:1993fg,Kosowsky:2001xp, Dolgov:2002ra,Nicolis:2003tg} which is enough for the point of this paper.
However, it certainly deserves improvement. We do not venture into this aspect in this work and just encourage that this question be re-examinated given the exciting experimental  prospects for GW detection from phase transitions we demonstrated in this analysis. 

\section*{Note added}

As this work was being completed, Ref.~\cite{Caprini:2006jb} appeared where they re-examined the calculation of the gravitational wave background from turbulence. They disagree with and correct the dispersion relation used for gravitational waves in~\cite{Kosowsky:2001xp,Dolgov:2002ra,Nicolis:2003tg}. This leads, in particular, to a different prediction for the peak frequency as well as a different spectral dependence.  Re-examination of the bubble collision spectrum is underway~\cite{CDS}. These new results can marginally affect the detectability regions of Figs.~\ref{fig:LISAcontours}, \ref{fig:BBO} and~\ref{fig:LIGOcontours} but the overall conclusions will remain the same.

\section*{Acknowledgments}

We are indebted to Alessandra Buonanno for very useful discussions. We also thank A.~Nicolis for a clarification on his work, M. Maggiore for pointing out some reference as well as C.~Caprini and R.~Durrer for discussions and finally James Wells for stimulating conversations at the early stage of this project.


\end{document}